\documentstyle[sprocl]{article}

\input{psfig}

\def\be{\begin{equation}}
\def\ee{\end{equation}}
\def\bea{\begin{eqnarray}}
\def\eea{\end{eqnarray}}

\begin{document}

\title{STRING COSMOLOGY AND CHAOS}

\author{J. D. BARROW}

\address{Astronomy Centre, University of Sussex, Falmer, Brighton,\\
 BN1 9QJ, United Kingdom\\ E-mail: J.D.Barrow@sussex.ac.uk} 

\author{M. P. D\c{A}BROWSKI \footnote{a presenter}}

\address{Astronomy Centre, University of Sussex, Falmer, Brighton,\\
 BN1 9QJ, United Kingdom\\ E-mail: mpd@star.cpes.susx.ac.uk \\and \\
Institute of Physics, University of Szczecin, Wielkopolska 15, Szczecin\\
70-451, Poland\\ E-mail: mpdabfz@uoo.univ.szczecin.pl}


\maketitle\abstracts{We investigate Bianchi type IX ''Mixmaster'' universes within the framework
of the low-energy tree-level effective action for string theory, which (when
the ''stringy'' 2-form axion potential vanishes) is formally the same as the
Brans-Dicke action with $\omega =-1$. We show that, unlike the case of
general relativity in vacuum, there is {\bf no Mixmaster chaos} in
these string cosmologies. In the Einstein frame an infinite sequence of
chaotic oscillations of the scale factors on approach to the initial
singularity is impossible, as it was in general relativistic Mixmaster
universes in the presence of stiff -fluid matter (or a massless scalar
field). A finite sequence of oscillations of the scale factors approximated
by Kasner metrics is possible, but it always ceases when all expansion rates
become positive. In the string frame the evolution through Kasner epochs
changes to a new form which reflects the duality symmetry of the theory.
Again, we show that chaotic oscillations must end after a finite time. The
need for {\bf duality symmetry} appears to be incompatible with the presence of
chaotic behaviour as $t\to 0$. We obtain our results using the Hamiltonian
qualitative cosmological picture for Mixmaster models. We mention possible
relations of this picture to diffeomorphism-independent methods of measuring
chaos in general relativity. Finally, we discuss the effect of
inhomogeneities and higher dimensions on the possible emergence of chaos
within the string cosmology.}

\section{Chaos in Relativity and Cosmology}

Chaotic systems arise in many areas of astronomy. Early studies of the
stability of the solar system by Poincar\'e have led to the modern
investigations of the motions of the satellites of Saturn and the chaotic
evolution of planetary obliquities by Laskar \cite{laskar}. The modern
resurgence of interest in chaotic dynamics was partly stimulated by the work
of H\'enon-Heiles on the orbits of stars in the disk of the Milky Way. These
systems have been studied within the context of Newton's theory and all the
classical criteria for defining chaos, in particular, non-vanishing Lyapunov
exponents and metric entropy can be applied. However, general relativistic
systems are different because they are automatically coordinate covariant
and are best studied using covariant measures of chaos since Lyapunov
exponents are explicitly time-dependent and also depend on the measure in
the phase space which makes them also spatial diffeomorphism dependent.

\subsection{Mixmaster Universe. Hamiltonian Approach}

\label{subsec:ham}

The paradigm for studies of general relativistic chaos in cosmology has been
the anisotropic Bianchi type IX universe or Mixmaster universe \cite{bkl}.
Different arguments for the appearance of chaos have been applied; the most
transparent is the Hamiltonian approach \cite{ryan} which reduces the
problem to the motion of a universe 'point' in a time-dependent,
steep-walled minisuperspace potential. These methods were applied to show
that a system describing the vacuum Bianchi IX model always allows the
universe point to catch up with the expanding potential walls and make
infinitely many bounces on the approach to the initial singularity. Then,
since every time the universe point is scattered the uncertainty in the
initial conditions grows, the system is chaotic. The Kasner indices fulfil
the requirements 
\begin{equation}
\sum_{i=1}^3p_i=1,\hspace{2.cm}\sum_{i=1}^3p_i^2=1.
\end{equation}
These conditions, in particular, do not allow the universe to enter the
regime in which it can be isotropic  (there is no closed vacuum Friedmann
universe). It was also found \cite{b9stiff} that the admission of
stiff-fluid matter with equation of state  $p=\varrho $ changes the Kasner
constraints to 
\begin{equation}
\ \sum_{i=1}^3p_i=1,\hspace{2.cm}\sum_{i=1}^3p_i^2=1-p_4^2,
\end{equation}
with $p_4$ playing the role of the fourth Kasner index - a constant which
come from the conservation law for the density. Such a choice admits a range
of the indices to include isotropic Friedmann model ($p_i=1/3$) and allows
the monotonic evolution of the three scale factors ($-1/3\leq p_1\leq
1/3,0\leq p_2\leq 2/3,1/3\leq p_3\leq 1$). In the Hamiltonian picture this
means that the universe point cannot be scattered against the potential
walls infinitely many times and after making a finite number of chaotic
oscillations the evolution becomes monotonic and chaos ceases. Positive
values of the indices correspond to the universe point having a normal
component of its velocity towards the wall which is lower than the velocity
of recession of the wall.

\subsection{Establishing Chaos}

\label{subsec:est}

Numerical studies of Bianchi IX universes confirmed its chaotic behaviour
although Lyapunov exponents were shown to be asymptotically tending to zero
through positive values in some cases \cite{rugh}. In view of their
non-covariant nature some other tools have been proposed. The first was to
introduce the so-called `Lyapunov-like exponents' \cite{szydlo} which can be
expressed in terms of Ricci scalar and other invariants of the Riemann
tensor. This characterisation relied on the reduction of Hamiltonian flows
to geodesic flows on some Riemann manifolds. The second was to apply other
invariant measures of chaos such as multifractal dimensions (information
dimension, correlation dimension) \cite{neil}. Both methods can show in a
coordinate independent way that the Mixmaster universe is chaotic.

\section{String Cosmology and Chaotic Oscillations}

\subsection{String (Pre-Big-Bang) Cosmology}

The low-energy effective-action for bosonic string theory was first
presented in \cite{action} and the field equations, up to the first order in
the inverse string tension $\alpha ^{\prime }$ in the \textit{string frame,}
are given by 
\begin{eqnarray}
R_\mu ^\nu +\nabla _\mu \nabla ^\nu \phi -\frac 14H_{\mu \alpha \beta
}H^{\nu \alpha \beta } &=&0,  \nonumber \\
R-\nabla _\mu \phi \nabla ^\mu \phi +2\nabla _\mu \nabla ^\mu \phi -\frac
1{12}H_{\mu \nu \beta }H^{\mu \nu \beta } &=&0, \\
\nabla _\mu \left( e^{-\phi }H^{\mu \nu \alpha }\right)  &=&0,  \nonumber
\end{eqnarray}
where $\phi $ is the dilaton field, $H_{\mu \nu \beta }=6\partial _{[\mu
}B_{\nu \beta ]}$ is the field strength of the antisymmetric tensor $B_{\mu
\nu }=-B_{\nu \mu }$ (usually called the axion). These equations with
vanishing axion are the same as those of Brans-Dicke scalar-tensor gravity
with a Brans-Dicke parameter $\omega =-1$. One recovers the Einstein limit
from (3) when the axion vanishes and dilaton is constant. The equations (3)
can be conformally transformed to the \textit{Einstein frame} \cite{chaos}
in which the field equations are those of Einstein relativity with two 
energy-momentum tensors for the dilaton (effectively stiff-fluid) and the
axion .

Isotropic cosmology based on the equations (3) is totally different from the
standard general relativistic cosmology since it admits a phase of expansion
for\textit{\ negative }times as well as for positive times (with the
singularity formally located at $t=0$). Because of this, it was originally
called \textit{pre-big-bang} cosmology \cite{pre}. The novel feature is that
it admits \textit{superinflation} which is driven by the kinetic energy of
the dilaton rather than by the potential energy of the inflaton, as it is in
the ordinary inflationary  picture. Superinflation (which is actually
power-law inflation), however, appears for negative times after which the
universe apparently approaches a stringy phase (strong curvature and
coupling regime) where the effective equations (3) break down, before
finally evolving towards a radiation-dominated expanding phase. Initially,
the universe is approximated by the perturbative string Minkowski vacuum.
This is different from standard inflationary picture in which the universe
tunnels quantum mechanically from ''nothing'', then undergoes inflation,
reheating and radiation-dominated expansion. The two types of the evolution
of the scale factor (superinflation and radiation-dominated evolution) are 
\textit{dual} to each other in the sense of string theory and this type of
symmetry (including time-reflection) transforms large values of the scale
factor at negative times into small values of the scale factor at  positive
times and vice versa. This feature is called scale factor duality \cite{pre}. Scale factor duality is an example of a more general continuous global $
O(d,d)$-symmetry of the string theory which is called $T$-duality \cite
{chaos}. 

Anisotropic cosmologies based on the equations (3) have also been studied.
This was done for universes of all Bianchi types and for Kantowski-Sachs.
Because of the homogeneity there was no problem in admitting a homogeneous
(i.e. spatially independent) dilaton field into all these geometries,
although this was not the case with axion. By analogy with the
electromagnetic field, one suspects that the admission of a
spatially-independent antisymmetric tensor potential $B_{\mu \nu }=B_{\mu
\nu }(t)$ should have elementary meaning (the so called 'elementary
ansatz'). However, as it was proven by many authors \cite{ansatz}, this
ansatz necessarily requires a distinguished direction in space and prevents
the universe from isotropization at late times in Bianchi type I and
Kantowski-Sachs models. On the other hand, the elementary ansatz is not
allowed in some Bianchi models. In particular, in Bianchi IX we have shown
this is even the case for its axisymmetric subcase \cite{chaos}. This led
some authors to employ another ansatz in which the antisymmetric tensor
field strength $H_{\mu \nu \rho }$ is time-dependent (the so called
'solitonic ansatz').This is admitted by Bianchi IX geometries \cite{chaos}.
Lastly, for spatially homogeneous models with non-abelian symmetry one deals
with the so-called non-abelian duality \cite{nonabel}.

\subsection{Is There Chaos?}

The differences between Einstein theory and the effective bosonic string
theory based on the equations (3) lead us to ask about the possible
emergence of chaotic behaviour for anisotropic Bianchi IX type models in
string theory. As mentioned in Section 1, the vacuum Bianchi IX model in
general relativity is chaotic while the stiff-fluid (scalar field with no
potential) model is not chaotic. In order to check whether there is chaos in
stringy Bianchi IX models one should apply one of the methods discussed in
Section 1.

It is easy to answer the question about chaos in the Einstein frame with
homogeneous dilaton and axion under the solitonic ansatz since both axion
and dilaton fields act as stiff fluids. One can show that the system
oscillates only finite number of times and then chaos ceases \cite{chaos}.
However, as most of the physics of string cosmology should refer to the
string frame, one needs to determine whether there are chaotic oscillations
in that frame. Of course, one might argue that the physics should be
frame-independent, but in the context of all changes of scale this is not
necessarily a trivial statement. In order to confirm this we applied the
Hamiltonian methods mentioned in Section 1.1. We calculated the velocity of
a universe particle $v_p$ moving against the potential walls and compared it
with the maximum apparent velocity of a wall in the axion frame $v_{max}$
(which is directly related to the string frame \cite{hamilt}) 
\begin{equation}
v_p=\sqrt{\psi _{+\tau }^2+\psi _{-\tau }^2}<v_{max}\approx \sqrt{\psi
_{+\tau }^2+\psi _{-\tau }^2+\frac 1{12}\phi _\tau ^2+\frac
1{12}A^2e^{-2\phi }},
\end{equation}
with $\psi _{+}$ and $\psi _{-}$ being the curvature anisotropies, subscript 
$\tau $ means the derivative with respect to a rescaled cosmic time $%
dt=d\tau e^{-\bar \phi }$ with $\bar \phi $ being the shifted dilaton \cite
{hamilt}. Here, $A$ is a constant determining the strength of the axion
field (using the solitonic ansatz). The relation (4) shows that, unless both
dilaton and axion vanish $\phi =A=0$ (which is the case for the vacuum
general relativity model) the universe particle cannot catch up with the
walls infinitely many times. The angle of incidence of the universe particle
moving towards a wall is too small sometimes and there are three regions in
the corners of the potential which are excluded from scatterings. As a
consequence, the universe makes a finite number of oscillations which stop
at some moment and then the scale factors all evolve monotonically towards a
curvature singularity \cite{chaos}.

There are some other interesting points. Firstly, the Kasner conditions are
different from those of vacuum or stiff-fluid case of Section 1, and read 
\begin{equation}
\sum_{i=1}^3p_i=1-p_4,\hspace{2.cm}\sum_{i=1}^3p_i^2=1,
\end{equation}
where the fourth Kasner index $p_4$ arises from the field conservation law.
This has consequences for the explicit duality-chaos relations which give
some insight into the impossibility of the co-existence of both in one
physical situation. There exist some chaotic Kasner-to-Kasner transitions
that are also duality-related transitions of the form 
\begin{equation}
p_1\to -p_1\hspace{0.3cm},p_2\to -p_2\hspace{0.3cm},p_3\to -p_3\hspace{0.3cm}%
,p_4\to -(p_4+2),
\end{equation}
which seem to prevent chaos. On the other hand, duality-related regions in
the parameter space (4) are given by 
\begin{eqnarray}
-1-\sqrt{3}\leq p_4\leq -1,-\frac 1{\sqrt{3}}\leq p_1\leq \sqrt{\frac 23}%
,-\frac 23\leq p_2\leq \frac 12\sqrt{\frac 23},-1\leq p_3\leq -\frac 12\sqrt{%
\frac 23};  \nonumber
\end{eqnarray}
and 
\begin{eqnarray}
-1\leq p_4\leq -1+\sqrt{3},-\sqrt{\frac 23}\leq p_1\leq \frac 1{\sqrt{3}%
},-\frac 12\sqrt{\frac 23}\leq p_2\leq \frac 23,\frac 12\sqrt{\frac 23}\leq
p_3\leq 1.  \nonumber
\end{eqnarray}
Of course, bearing in mind the behaviour of the system in the Einstein
frame, one can argue that this is just because we are dealing with the
stiff-fluid. Secondly, as one can learn from the field equations with
explicit Bianchi IX geometry \cite{chaos}, that the dilaton field appears as
a factor in front of the scale factors on the right-hand side of these
equations, i.e., 
\begin{eqnarray}
a^4e^{-2\phi } &\propto &t^{(2p_1-p_4)}=t^{(1+p_1-p_2-p_3)},  \nonumber \\
b^4e^{-2\phi } &\propto &t^{(2p_2-p_4)}=t^{(1+p_2-p_3-p_1)}, \\
c^4e^{-2\phi } &\propto &t^{(2p_3-p_4)}=t^{(1+p_3-p_1-p_2)},  \nonumber
\end{eqnarray}
where $a,b,c$ are the scale factors and $t$ is the cosmic (comoving proper)
time. This even restricts the possibilities for chaos to begin, as occurs
for instance in five-dimensional vacuum Bianchi IX models \cite{5dim}. The
admissible values of the first two Kasner indices for chaotic oscillations
to begin are given in Fig. 1. As one can easily see, the (dual) isotropic
Friedmann cases ($p_4=-1+\sqrt{3},p_1=p_2=p_3=1/\sqrt{3}$ and $p_4=-1-\sqrt{3%
},$ $p_1=p_2=p_3=-1/\sqrt{3}$) are excluded.

Figure 1: \hfill

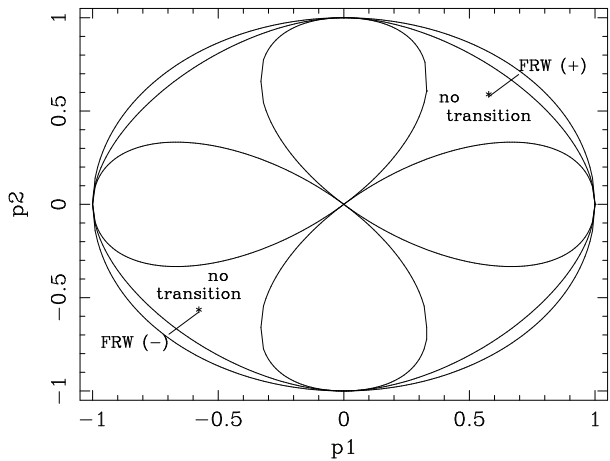
\centerline{\psfig{figure=dabrfig.ps,height=8cm} \hskip 2cm}


\section{Perspectives}

One of the simplest ways to recover chaos in low-energy string cosmology
would be by the introduction of the dilaton potential (cosmological term)
with possibly similar effect as the mass term in general relativistic scalar
field cosmologies. This is also a situation which breaks duality.

Another way might be to appeal to higher dimensions or inhomogeneities. It
is well-known that anisotropic 5-dimensional vacuum models are not chaotic 
\cite{5dim}. This can be seen by analysing the generalized admissible ranges
of Kasner indices 
\begin{equation}
\sum_{i=1}^4p_i=1,\hspace{2.cm}\sum_{i=1}^4p_i^2=1
\end{equation}
similarly to our Fig.1 and by proving that the extra dimension slows the
universe point which cannot catch up with the walls, finally leading to the
monotonic (non-chaotic) expansion. If one enlarges the number of dimensions
and also admits inhomogeneities or off-diagonal metric elements to the
homogeneous diagonal Bianchi IX models, then the conditions (6) generalize
to become 
\begin{equation}
\sum_{i=1}^{D-1}p_i(x)=1,\hspace{2.cm}\sum_{i=1}^{D-1}p_i^2(x)=1,
\end{equation}
where $D$ is the number of spacetime dimensions and $x$ are spatial
coordinates \cite{multidim}. The result, in the Hamiltonian picture, is that
the potential walls start rotating with the velocity of rotation going to
zero on the approach to singularity and the universe point can actually be
scattered infinitely many times, but this does not happen if the number of
spacetime dimensions satisfies $D\ge 11$. The same effect of recovering
chaos in multidimensional models appears in non-diagonal homogeneous Bianchi
IX  models of $D\le 10$ \cite{demaret}. 

These results show that the dimensional structure of type IX universes is
non-trivial. The $D=4$ case is the most complicated possible. Chaos occurs
in diagonal vacuum models and is created by the intrinsically general
relativistic parts of the Weyl curvature anisotropy which dominate the
dynamics infinitely often on approach to a singularity, causing bounces.
Physically, the motion of gravitational waves on a simple background space
is curving up the space in the direction of propagation until the curvature
back reaction reverses is motion. Note that for most of the time the
3-curvature of the type IX universe is actually negative. As the number of
dimensions increases the situation becomes simpler. Off diagonal or
inhomogeneous contributions to the metric are required to create chaos in
the range $4\leq D\leq 10$, but when the dimensionality exceeds this the
evolution is no longer dominated by the intrinsically general relativistic
Weyl curvature effects: the Newtonian parts of the gravitational field
dominate on approach to the singularity. We should also note that one must
distinguish between Mixmaster models with full $SO(D-1)$ invariance and
those with a product manifold structure which behave effectively as if they
have fewer dimensions.  These features suggest further investigations since
superstring theories are formulated in $D=10$ spacetime dimensions, while $M$%
-theory with its low-energy supergravity limit is formulated in $D=11$
dimensions. Unfortunately, so far there appears to be no specific link
between the causes of the disappearance of chaos in $D>10$ general
relativity and the finiteness conditions that pick out $D=10$ string
theories. The disappearance of chaos in $D=11$ dimensions might suggest that 
$M$-theory has a simpler structure in this particular respect. However, this
might not necessarily be the case since its subsystems include Yang-Mills
fields which are chaotic because of the colour charges even in axisymmetric
Bianchi type I geometries \cite{Mtheory} but this chaos (unlike the
Mixmaster phenomenon) is not general relativistic in origin. This suggests
that we include other string modes together with a range of compactification
schemes \cite{hosoya} in order to answer the question about the existence of
chaos in string cosmology and its generalizations.

\section*{Acknowledgments}

MPD acknowledges the support of NATO/Royal Society and the Polish Research
Committee (KBN) under the grant No 2PO3B 096 10. JDB is supported by a PPARC
Senior Fellowship.

\section*{Appendix}


\section{References}

\begin{thebibliography}{99}
\bibitem{laskar}  J. Laskar and P. Robutel, Nature, \textbf{361}, 608 (1993);\\
J. Laskar \textit{et al.}, Nature, \textbf{361}, 615 (1993).
                
\bibitem{bkl}  V.A. Belinskii \textit{et al}, Sov. Phys. Uspekhi \textbf{102}%
, 745 (1971);\\ J.D. Barrow, \textit{Phys. Rev. Lett.} \textbf{46}, 963
(1981);\\ J.D. Barrow, \textit{Phys. Rep.} \textbf{85}, 1 (1982);\\ J.D.
Barrow, Gen. Rel. Gravitation, \textbf{14}, 1523 (1982);\\ D. Chernoff and
J.D. Barrow, Phys. Rev. Lett. \textbf{50}, 134 (1983);\\ J.D. Barrow, in 
\textit{Classical General Relativity, }eds. W. Bonnor, J. Islam and M.A.H.
MacCallum, (Cambridge University Press, Cambridge, 1984), pp. 25-41..

\bibitem{ryan}  M.P. Ryan and L.C. Shepley, \textit{Homogeneous Relativistic
Cosmologies} (Princeton University Press, Princeton, 1975).

\bibitem{szydlo}  M. Szyd{\l }owski, \textit{Phys. Lett.} A\textbf{176}, 22
(1993).

\bibitem{neil}  N.J. Cornish and J.J. Levin, \textit{Phys. Rev. Lett.} 
\textbf{78}, 998 (1997); {\emph{Phys. Rev.} D} \textbf{55}, 7489 (1997).

\bibitem{b9stiff}  V.A. Belinskii and I.M. Khalatnikov, \textit{Sov. Phys.
JETP} \textbf{36}, 591 (1973).

\bibitem{rugh}  B.J.T. Jones and S. Rugh, \textit{Phys. Lett.} A \textbf{147,%
}, 353 (1990).

\bibitem{action}  E.S. Fradkin and A.A. Tseytlin, {\emph{Nucl. Phys.} B} 
\textbf{267}, 1 (1985).

\bibitem{chaos}  J.D. Barrow and M.P. D\c abrowski, {\emph{Phys. Rev.} D} 
\textbf{57}, (1998).

\bibitem{pre}  G. Veneziano, {\emph{Phys. Lett.} B} \textbf{265}, 287 (1991);%
\\ M. Gasperini and G. Veneziano, {\emph{Phys. Rev.} D} \textbf{50}, 2519
(1994).

\bibitem{ansatz}  E.J. Copeland \textit{et al}, {\emph{Phys. Rev.} D} 
\textbf{51}, 1569 (1995);\\ J.D. Barrow and K.E. Kunze, {\emph{Phys. Rev.} D}
\textbf{55}, 623 (1997);\\ J.D. Barrow and M.P. D\c abrowski,{\emph{Phys.
Rev.} D} \textbf{55}, 630 (1997).

\bibitem{nonabel}  M. Gasperini \textit{et al}, {\emph{Phys. Lett.} B} 
\textbf{319}, 438 (1993).

\bibitem{hamilt}  R. Easther \textit{et al}, {\emph{Phys. Rev.} D} \textbf{53%
}, 4247 (1996);\\ M. Gasperini and G. Veneziano, \textit{Gen. Rel. Grav.} 
\textbf{28}, 1301 (1996).

\bibitem{dynsys}  J. Wainwright and L. Hsu,\textit{\ Class. Quantum Grav.}, 
\textbf{6}, 1409 (1989).

\bibitem{5dim}  J.D. Barrow and J. Stein Schabes,\textit{\ Phys. Rev.} D 
\textbf{32, }1595 (1985);\\ P. Halpern, {\emph{Phys. Rev.} D} \textbf{33},
354 (1986);\\ H. Ishihara, \textit{Prog. Theor. Phys}. \textbf{74}, 354
(1986).

\bibitem{hosoya}  T. Furosawa and A. Hosoya,\textit{\ Prog. Theor. Phys}. 
\textbf{73}, 467 (1985).

\bibitem{multidim}  V.A. Belinskii \textit{et al}, \textit{Adv. Phys.} 
\textbf{31}, 639 (1982);\\ J. Demaret \textit{et al}, {\emph{Phys. Lett.} B} 
\textbf{164}, 27 (1985);\\ J. Demaret \textit{et al}, {\emph{Phys. Lett.} B} 
\textbf{175}, 129 (1986);\\ A. Hosoya \textit{et al}, {\emph{Nucl. Phys.} B} 
\textbf{283}, 657 (1987).

\bibitem{demaret}  J. Demaret \textit{et al}, {\emph{Phys. Lett.} B} \textbf{%
211}, 37 (1988).

\bibitem{Mtheory}  D.V. Gal'tsov and M.S. Volkov, {\emph{Phys. Lett.} B} 
\textbf{256}, 17 (1991);\\ I. Ya. Aref'eva, P.B. Medvedev, O.A. Rytchkov and
I.V. Volovich, `Chaos in M(atrix) Theory', e-Print:hep-th/9710032;\\ 
B.K. Darian and H.P. K\"unzle, Class. Quant. Grav. \textbf{12}, 2651 (1995);\\J.D.
Barrow and J. Levin, \textit{Phys. Rev. Lett.} \textbf{80}, 656 (1998).
\end{thebibliography}
\end{document}